\RequirePackage[2020-02-02]{latexrelease}
\documentclass[onecolumn,10pt,aps]{revtex4}
\usepackage{amssymb}
\usepackage{graphicx}
\usepackage[cp1250]{inputenc}
\usepackage{amsmath}
\usepackage{amsfonts}
\usepackage{graphicx}
\usepackage{yfonts}
\usepackage{color}
\usepackage[normalem]{ulem}
\usepackage{amsthm}
\usepackage{bm}j
\usepackage{bbm}
\usepackage{subcaption}
\usepackage{float}
\usepackage{enumitem}

\DeclareMathOperator{\Tr}{Tr}

\def\<{\langle}
\def\>{\rangle}

\usepackage{placeins}
\def\oper{{\mathchoice{\rm 1\mskip-4mu l}{\rm 1\mskip-4mu l}
{\rm 1\mskip-4.5mu l}{\rm 1\mskip-5mu l}}}

\begin{document}

\title{Controlling nonlocality of bipartite qubit states via quantum channels}

\author{Adam Rutkowski${}^1$\footnote{e-mail:adam.rutkowski@ug.edu.pl} and Katarzyna Siudzi\'{n}ska${}^2$}

\affiliation{ ${}^1$Institute of Theoretical Physics and Astrophysics \\ Faculty of Mathematics, Physics and Informatics, University of Gda\'{n}sk, 80-952 Gda\'{n}sk, Poland\\
${}^2$Institute of Physics, Faculty of Physics, Astronomy and Informatics,\\ Nicolaus Copernicus University in Toru\'{n}, ul. Grudzi\c{a}dzka 5, 87-100 Toru\'{n}, Poland}

\begin{abstract}
In this paper, we analyze quantum channels derived from a class of two-qubit states known as the X states. In particular, we consider X states that break the Bell's CHSH condition and then characterize the associated inverse Choi-Jamio{\l}kowski maps, which we call nonlocality generating. Interestingly, the region corresponding to nonlocality generating channels shrinks as their stationary state departs from the maximally mixed state. Finally, we demonstrate special cases, where nonlocality can be fully characterized via as little as a single parameter. This approach significantly simplifies the analysis of nonlocality for the X states.
\end{abstract}

\maketitle

\section{Introduction}

Nonlocality is an important concept that allows us to distinguish between a quantum theory and local hidden variable models. Quantum states that are nonlocal violate Bell-like inequalities, such as the CHSH (Clauser-Horne-Shimony-Holt) inequality \cite{Clauser1969}, which provide a specific testable criteria. However, a complete parametrization of all local states have been found only for bipartite qubit systems \cite{HHH,Masanes}. For any other qudit states, the conditions for nonlocality are either necessary or sufficient and depend on the examined states \cite{Fonseca,Chen,Zhang,Wang}.

In the present study, we consider the inverse Choi-Jamio{\l}kowski isomorphism for nonlocal X states \cite{Eberly}, which are two-qubit circulant states \cite{Kossakowski}. The construction of circulant states is based on a direct sum decomposition of the Hilbert space with a circular structure. Popular special classes include maximally entangled states, positive partial transpose states \cite{PPT}, Werner states \cite{Werner}, isotropic states \cite{iso}, as well as the Choi matrix of the Choi map \cite{Choi_map} and its extensions.
Through the state-channel correspondence, a nonlocal bipartite state can be associated with a quantum channel (completely positive, trace-preserving map), which we call \textit{nonlocality generating}. The images of states and maps are isomorphic; however, this does not imply that the level of difficulty in problem-solving is the same. We show that, using quantum channels, one can more easily discern various properties, such as the minimal number of parameters required to control nonlocality. Many important channels belong to this class, including the Pauli channels and non-unital phase-covariant channels. It turns out that breaking locality can be controlled via as little as a single channel parameter.

This work focuses on the channels isomorphic to two-qubit states because there are no general necessary and sufficient nonlocality conditions for two-qudit states \cite{Brunner,Scarani}. It is well-established that if a state is nonlocal, then it is also entangled, but the inverse relation does not hold. Furthermore, in ref. \cite{Mendonca2014}, it is shown that, for every two-qubit state, there exists the corresponding X state that shares its same spectrum and entanglement properties. This includes measures such as concurrence, negativity, and relative entropy of entanglement. The X states frequently emerge in experimental setups, offering a practical framework for analyzing quantum operations, such as teleportation and quantum key distribution \cite{Nagaoka}. Their significance lies in bridging theoretical insights and experimental realizability, serving as a foundational model for quantum technologies \cite{Kossakowski}.

This article is organized as follows. Section II introduces the notion of circulant states and shows that a wide class of quantum qubit maps is isomorphic to two-qubit X states. In Section III, we recall the CHSH Bell's inequality and its relation to quantum nonlocality. Analytical conditions for the X states to break the CHSH condition are derived in terms of the channel parameters. The region of nonlocality generating qubit channels is analyzed for both unital and non-unital channels. Finally, in Section IV, we further analyze the property of nonlocality generation for important families of bistochastic (Pauli) channels and non-unital phase-covariant channels. We summarize our results in the Conclusions, where we also provide a list of open problems and further research.

\section{Quantum channels from circulant states}

\subsection{Circulant states}

Following the construction methods proposed in refs. \cite{Kossakowski,Adam}, let us define the basic $n$-dimensional subspace of $\mathbb{C}^{n}\otimes\mathbb{C}^{m}$ via
\begin{equation}
\Sigma_{0}=\text{{span}}_{\mathbb{C}}\left\{ \left|e_{0}\otimes f_{0}\right\rangle ,\:\left|e_{1}\otimes f_{1}\right\rangle ,\ldots,\left|e_{n-1}\otimes f_{n-1}\right\rangle \right\} ,
\end{equation}
where $\left|e_{i}\right\rangle$ and $\left|f_{i}\right\rangle$ are the basis vectors in $\mathbb{C}^{n}$ and $\mathbb{C}^{m}$, respectively (index addition mod $m$). 
Now, for any permutation $\pi$ from the symmetric group $\mathcal{S}_{m}$, we introduce the subspace
\begin{equation}
\Sigma_{0}^{\pi}=\text{{span}}_{\mathbb{C}}\left\{ \left|e_{0}\otimes f_{\pi\left(0\right)}\right\rangle ,\:\left|e_{1}\otimes f_{\pi\left(1\right)}\right\rangle ,\ldots,\left|e_{n-1}\otimes f_{\pi\left(n-1\right)}\right\rangle \right\} .
\end{equation}
Using the matrix representation $\Pi$ of the permutation operation $\pi$, we find that the relation between those subspaces reads $\Sigma_{0}^{\pi}=\left(\mathbb{I}_n\otimes\Pi\right)\Sigma_{0}$,
where $\mathbb{I}_n$ is the identity operator on $\mathbb{C}^n$. From now on, we restrict our considerations to such permutations that $\pi\left(0\right)=0$. This effectively means that $\left|e_{0}\otimes f_{0}\right\rangle$ always belongs to the subspace $\Sigma_{0}$ in this decomposition. Since the total Hilbert space dimension is $nm$, we need to define $m-1$ additional subspaces
\begin{equation}
\Sigma_{i}^{\pi}=\left(\mathbb{I}\otimes S^{i}\right)\Sigma_{0}^{\pi}=\left(\mathbb{I}\otimes S^{i}\Pi\right)\Sigma_{0},\qquad i=1,\ldots,m-1.
\end{equation}
In the above formula, $S$ is the cyclic matrix corresponding to the shift operator in $\mathbb{C}^{m}$; i.e., $S\left|e_{i}\right\rangle =\left|e_{i+1}\right\rangle$. On every subspace $\Sigma_{i}^{\pi}$, we define $m$ operators
\begin{equation}
\mathcal{O}_{\alpha}^{\pi}=\sum_{i,j=0}^{m-1}a_{ij}^{\left(\alpha\right)}e_{ij}\otimes S^{\alpha}f_{\pi\left(i\right),\pi\left(j\right)} (S^{\alpha})^T,
\end{equation}
where $e_{ij}=|e_i\>\<e_j|$,
$f_{\pi(i),\pi(j)}=\left|f_{\pi(i)}\left\rangle \right\langle f_{\pi(j)}\right|$, and $a_{ij}^{\left(\alpha\right)}$ are complex numbers. Finally, it can be easily shown that a direct sum over all $\Sigma_{i}^{\pi}$ spans the entire space $\mathbb{C}^{n}\otimes\mathbb{C}^{m}$,
\begin{equation}
\Sigma_{0}^{\pi}\oplus\Sigma_{1}^{\pi}\oplus\ldots\oplus\Sigma_{m-1}^{\pi}=\mathbb{C}^{n}\otimes\mathbb{C}^{m}.
\end{equation}
Therefore, the circulant operators corresponding to this distinguished decomposition are constructed as follows,
\begin{equation}\label{Ostancykliczny}
\mathcal{O}_{\pi}=\sum_{\alpha=0}^{m-1}\mathcal{O}_{\alpha}^{\pi}.
\end{equation}
Note that in the special case where $n=m=2$, there exists a unique decomposition into the direct sum of
\begin{equation}
\begin{array}{c}
\Sigma_{0}=\text{span}_{\mathbb{C}}\left\{ e_{0}\otimes f_{0},\:e_{1}\otimes f_{1}\right\}, \\
\Sigma_{1}=\text{span}_{\mathbb{C}}\left\{ e_{0}\otimes f_{1},\:e_{1}\otimes f_{0}\right\}.
\end{array}
\end{equation}

Let us now introduce the notion of circulant states in the context of a physical application of circulant operators. By demanding that $\mathcal{O}{\pi}:\mathbb{C}^{nm}\to\mathbb{C}^{nm}$, $\mathcal{O}{\pi}\geq 0$, and $\Tr(\mathcal{O}_{\pi})=1$, one defines {\it circulant states}. 
 They were first studied by Eberly and Yu \cite{Eberly,Eberly2} for bipartite qubit systems under the name {\it X states} due to the form of their matrix representation in the operational basis 
$\{|k\>\otimes|\ell\>;\,k,\ell=0,1\}$,
\begin{equation}\label{X}
\rho=
\left(\begin{array}{cc|cc}
a & 0 & 0 & w \\
0 & b & z & 0 \\
\hline 0 & z^* & c & 0 \\
w^* & 0 & 0 & d
\end{array}\right).
\end{equation}
For $\rho$ to be a quantum state ($\rho\geq 0$, $\text{Tr}(\rho)=1$), the matrix parameters are constrained via
\begin{equation}\label{pos}
\sqrt{ad}\geq |w|,\qquad \sqrt{bc}\geq |z|,
\end{equation}
$d=1-a-b-c$, and $a,b,c,d\geq 0$. The X states are crucial in quantum physics due to their simple structure and broad applicability in quantum information theory. Their sparse density matrix allows for efficient computation of quantum correlations, making them ideal for studying entanglement dynamics and decoherence in open quantum systems \cite{Eberly}. Additionally, X states frequently emerge in experimental setups, offering a practical framework for analyzing quantum operations, such as teleportation and quantum key distribution \cite{Nagaoka}. Their significance lies in bridging theoretical insights and experimental realizability, serving as a foundational model for quantum technologies \cite{Kossakowski}.
As mentioned earlier, for every two-qubit state, there exists the corresponding X state that shares the same spectrum and entanglement properties as that state \cite{Mendonca2014}. The authors provide a parametric form of a unitary transformation that maps arbitrary two-qubit states to their X state counterparts while preserving entanglement. This shows that X states are sufficient to describe the entanglement of any two-qubit system.

\subsection{Qubit channels}
In this section, we  introduce a method for describing quantum channels. We  also establish a clear correspondence between a quantum channel and a quantum state and explain how the conditions for being a quantum state transfer to quantum channels.
Consider the most general trace-preserving qubit map $\Phi=\mathcal{U}\Lambda\mathcal{V}$ \cite{Szarek}, where $\mathcal{U}$ and $\mathcal{V}$ are arbitrary unitary rotations and
\begin{equation}
\label{GeneralLambda}
\begin{split}
\Lambda[X]=\frac 12 \Big[&(\mathbb{I}+t_1\sigma_1+t_2\sigma_2+t_3\sigma_3)\Tr(X)\\&
+\lambda_1\sigma_1\Tr(X\sigma_1)+\lambda_2\sigma_2\Tr(X\sigma_2)
+\lambda_3\sigma_3\Tr(X\sigma_3)\Big].
\end{split}
\end{equation}
In the above formula, $\lambda_k$, $k=1,2,3$, are the eigenvalues of $\Lambda$ to the eigenvectors $\sigma_k$, which are the Pauli matrices. The remaining parameters $t_k$, $k=1,2,3$, determine the stationary state of the map,
\begin{equation}
    X_\ast=\Lambda[X_\ast]=\frac 12 \left[\mathbb{I}+\sum_{k=1}^3\frac{t_k}{1-\lambda_k}\sigma_k\right],
\end{equation}
which ranges from pure (e.g., $|0\>\<0|$ and $|1\>\<1|$ for $t_1=t_2=0$, $t_3=\pm(1-\lambda_3)$) to maximally mixed $\mathbb{I}/2$ (for $t_1=t_2=t_3=0$).

Every qudit channel $\Lambda$ is isomorphic to a two-qudit state $\rho_\Lambda$ via the celebrated Choi-Jamio{\l}kowski isomorphism \cite{Choi,Jamiolkowski}. 
This isomorphism is crucial in quantum information and physics because it simplifies the study of quantum channels. It allows quantum maps to be represented as single operators (Choi matrix), making their mathematical and experimental analysis more accessible. Furthermore, it plays a central role in quantum process tomography, where reconstructing a quantum channel experimentally often involves estimating its Choi matrix. By linking quantum operations with entangled states, the Choi-Jamio{\l}kowski isomorphism also facilitates the study of entanglement and nonlocality, offering valuable insights into the behavior of quantum systems.

For qubit channels, we calculate
\[
\rho_\Lambda=(\oper\otimes\Lambda)[P_+]
\]
by acting with the extension of $\Lambda$ on the maximally entangled state $P_+=(1/4)\sum_{k,\ell=0}^1|k\>\<\ell|\otimes|k\>\<\ell|$. 
Its Choi matrix representation reads
\begin{equation}\label{rho}
\rho_\Lambda=\frac 14
\left(\begin{array} {cc|cc}
1+\lambda_3+t_3 & t_1-it_2 & 0 & \lambda_1+\lambda_2 \\
t_1+it_2 & 1-\lambda_3-t_3 & \lambda_1-\lambda_2 & 0 \\
\hline
0 & \lambda_1-\lambda_2 & 1-\lambda_3+t_3 & t_1-it_2 \\
\lambda_1+\lambda_2 & 0 & t_1+it_2 & 1+\lambda_3-t_3
\end{array}\right).
\end{equation}
Observe that $\rho_\Lambda$ has the structure of an X state if and only if $t_1=t_2=0$. 
By comparing eq. (\ref{rho}) with the general form of an X state in eq. (\ref{X}), we identify
\begin{equation*}
z=\frac{\lambda_1-\lambda_2}{4},\qquad w=\frac{\lambda_1+\lambda_2}{4},\qquad a=\frac{1+\lambda_3+t_3}{4},
\end{equation*}
\begin{equation*}
b=\frac{1-\lambda_3-t_3}{4},\qquad c=\frac{1-\lambda_3+t_3}{4},\qquad d=\frac{1+\lambda_3-t_3}{4}.
\end{equation*}
The positivity conditions for the state $\rho_\Lambda$ in eq. (\ref{pos}) are equivalent to the complete positivity conditions for the channel $\Lambda$. Therefore, $\Lambda$ is completely positive if and only if the following two inequalities are satisfied,
\begin{equation}\label{CPcon}
    \sqrt{(1\pm\lambda_3)^2-t_3^2}\geq|\lambda_1\pm\lambda_2|.
\end{equation}
This agrees with the generalized Fujiwara-Algoet conditions  \cite{Fujiwara2}. Note that eq. (\ref{CPcon}) is symmetric with respect to the change of sign of the parameter $t_3$.

\section{Breaking the CHSH condition}

The CHSH inequality \cite{Clauser1969} (Clauser-Horne-Shimony-Holt) is a form of Bell's inequality, formulated in 1969, and is a key tool in studying the discrepancies between local hidden variable theories and quantum mechanics. It is primarily used in the context of quantum nonlocality tests, demonstrating that certain quantum phenomena cannot be explained by classical theories, which assume local realism.

Consider two spatially separated systems (e.g., particles) and two observers, A and B, who perform measurements on their systems. Each observer can choose between two possible detector settings (for A, the settings are \( A_1 \) and \( A_2 \), and for B, \( B_1 \) and \( B_2 \)). The measurement results can take values of either $+1$ or $-1$. The functional form of the CHSH inequality is based on the expected value of the product of measurement outcomes for different detector settings,
\[
S = |E(A_1, B_1) + E(A_1, B_2) + E(A_2, B_1) - E(A_2, B_2)|,
\]
where \( E(A_i, B_j) \) denotes the expected value of the product of Alice's and Bob's measurement results for settings \( A_i \) and \( B_j \).
In classical local realism, the value of \( S \) satisfies the inequality \cite{Clauser1969}
  \[
  S \leq 2.
  \]
In quantum theory, it is possible to achieve a larger value, with a maximum of
  \[
  S \leq 2\sqrt{2} \approx 2.828.
  \]
Quantum experiments, such as those involving entangled photons, have repeatedly shown that the value of \( S \) exceeds the classical bound of 2, providing strong evidence of quantum nonlocality. This result implies that quantum mechanics violates classical local realism and cannot be explained by hidden variable theories.
The CHSH inequality is fundamental in tests of quantum nonlocality, showing that quantum mechanics is incompatible with the assumption of local realism. This has profound implications for our understanding of the nature of reality and practical applications in quantum cryptography and quantum communication.

In ref. \cite{HHH}, the Horodecki family demonstrate that the CHSH condition for qubit pairs can be equivalently expressed in terms of the correlation matrix $T_{jk}=\Tr\left(\rho\sigma_j\otimes\sigma_k\right)$, $j,k=1,2,3$. For the X state in eq. (\ref{X}), one has
\begin{equation}
T=
\left(\begin{array} {cc cc}
w+w^*+z+z^* & i(w-w^*-z+z^*) & 0 \\
-i(-w+w^*-z+z^*) & -w-w^*+z+z^* & 0 \\
0 & 0 & a-b-c+d
\end{array}\right).
\end{equation}
Note that the rows and columns associated with $\sigma_0=\mathbb{I}$ is omitted due to providing no information about the qubit correlations. 
Let us calculate \(T^\dagger T \). In our case,
\begin{equation}
T^\dagger T=
\left(\begin{array} {cc cc}
4|z+w^*|^2 & 4i(wz-w^*z^*) & 0 \\
4i(wz-w^*z^*) & 4|z-w^*|^2 & 0 \\
0 & 0 & (a-b-c+d)^2
\end{array}\right),
\end{equation}
and its eigenvalues read
\begin{equation}
s_1=(a-b-c+d)^2,\qquad
s_2=4(|w|-|z|)^2,\qquad
s_3=4(|w|+|z|)^2.
\end{equation}
According to the Horodecki criterion \cite{HHH}, the X state is nonlocal (breaks the CHSH condition) if and only if
\begin{equation}
\max\{s_1+s_2,s_2+s_3,s_3+s_1\}>1.
\end{equation}
For the Choi-Jamio{\l}kowski state $\rho_\Lambda$, we recover
\begin{equation}
s_1=\lambda_3^2,\qquad
s_2=\frac 14(|\lambda_1+\lambda_2|-|\lambda_1-\lambda_2|)^2,\qquad
s_3=\frac 14(|\lambda_1+\lambda_2|+|\lambda_1-\lambda_2|)^2.
\end{equation}
It is straightforward to show that $s_3\geq s_2$ always holds, so that the Horodecki condition reduces to
\begin{equation}\label{CH1}
\begin{cases}
&|\lambda_1+\lambda_2|+|\lambda_1-\lambda_2|<2\sqrt{1-\lambda_3^2},\\
&(|\lambda_1+\lambda_2|-|\lambda_1-\lambda_2|)^2\leq 4\lambda_3^2,
\end{cases}
\end{equation}
or
\begin{equation}\label{CH2}
\begin{cases}
&|\lambda_1+\lambda_2|<2,\\
&(|\lambda_1+\lambda_2|-|\lambda_1-\lambda_2|)^2>4\lambda_3^2.
\end{cases}
\end{equation}
In particular, conditions (\ref{CH1}) and (\ref{CH2}) correspond to $s_1+s_3>1$ and $s_2+s_3>1$, respectively. Note that there is no dependence on the value of $t_3$, only on the eigenvalues $\lambda_1,\lambda_2,\lambda_3$ of the qubit channel. From its complete positivity conditions, it also follows that the first inequality in the second set of conditions is always satisfied provided that $\lambda_3<1$. In particular, the identity map $\oper$ satisfies neither of the eqs. (\ref{CH1}--\ref{CH2}). Hence, there exist no dynamical maps (time-parameterized families of quantum channels) $\{\Lambda(t);\,t\geq 0,\,\Lambda(0)=\oper\}$ that are generating nonlocality at all times $t\geq 0$. However, they can eventually break the CHSH conditions or never break them at all.

The ranges of channel eigenvalues $\lambda_k$ corresponding to nonlocality generating channels are plotted in Fig. 1 for different values of $t_3$. Observe that the red and blue shapes are clearly separated. Also, there is a symmetry plane at $\lambda_1=\lambda_2$. For $t_3=0$, the complete positivity region is a tetrahedron. Therefore, it is clearly visible that the Choi-Jamio{\l}kowski states $\rho_\Lambda$ break the CHSH condition for the vast majority of unital qubit channels $\Lambda$. Then, with the increase of $t_3$, the plotted region becomes smaller as its edges are cut out. For some $1/4<t_3\leq 1/2$, its shape smoothens (the set becomes convex) and remains mostly the same while shrinking with a further increase of $t_3$. Finally, at $t_3=1$, the region vanishes because then the only completely positive map is the maximally depolarizing channel $\Phi_0[X]=\mathbb{I}\Tr(X)/2$, for which the nonlocality conditions are not obeyed.

    \begin{figure}[H]
    \centering
 \begin{subfigure}[b]{0.4\linewidth}
        \centering
        \includegraphics[width=0.8\linewidth]{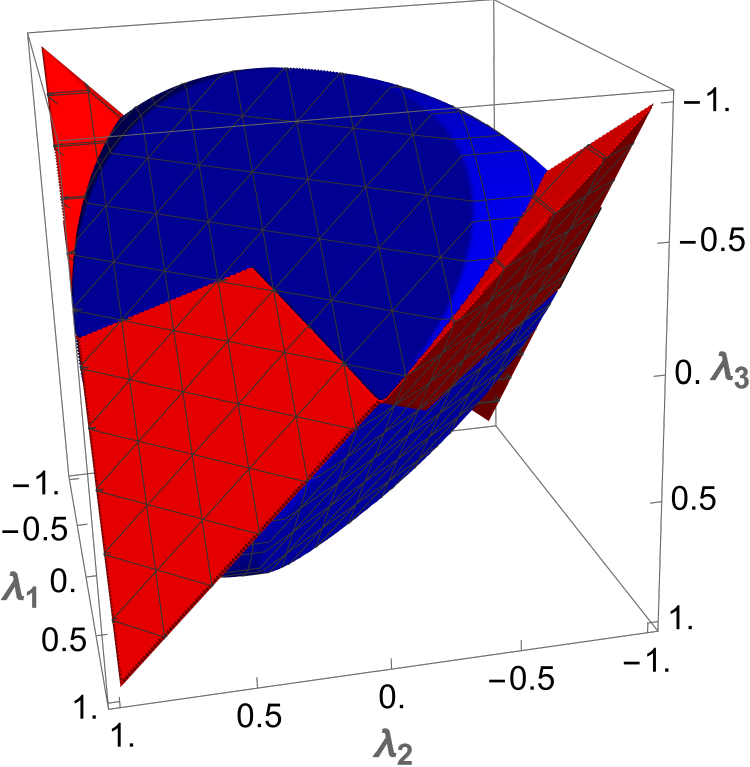}
        \caption{\(t_3 = 0\)}
    \end{subfigure}
    \qquad
    \begin{subfigure}[b]{0.4\linewidth}
        \centering
        \includegraphics[width=0.8\linewidth]{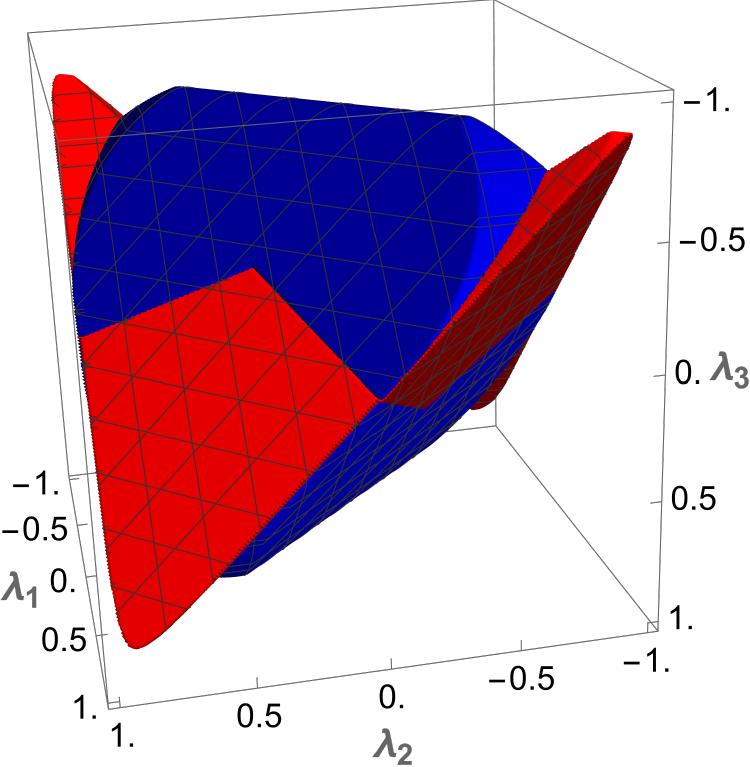}
        \caption{\(t_3 = 1/8\)}
    \end{subfigure}

    \vskip\baselineskip
    
 \begin{subfigure}[b]{0.4\linewidth}
        \centering
        \includegraphics[width=0.8\linewidth]{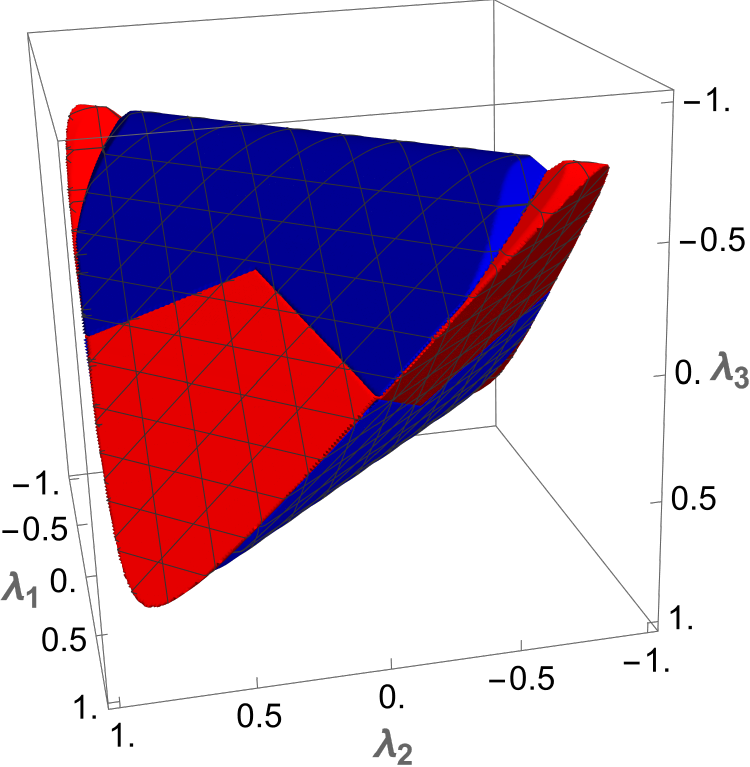}
        \caption{\(t_3 = 1/4\)}
    \end{subfigure}
    \qquad
    \begin{subfigure}[b]{0.4\linewidth}
        \centering
        \includegraphics[width=0.8\linewidth]{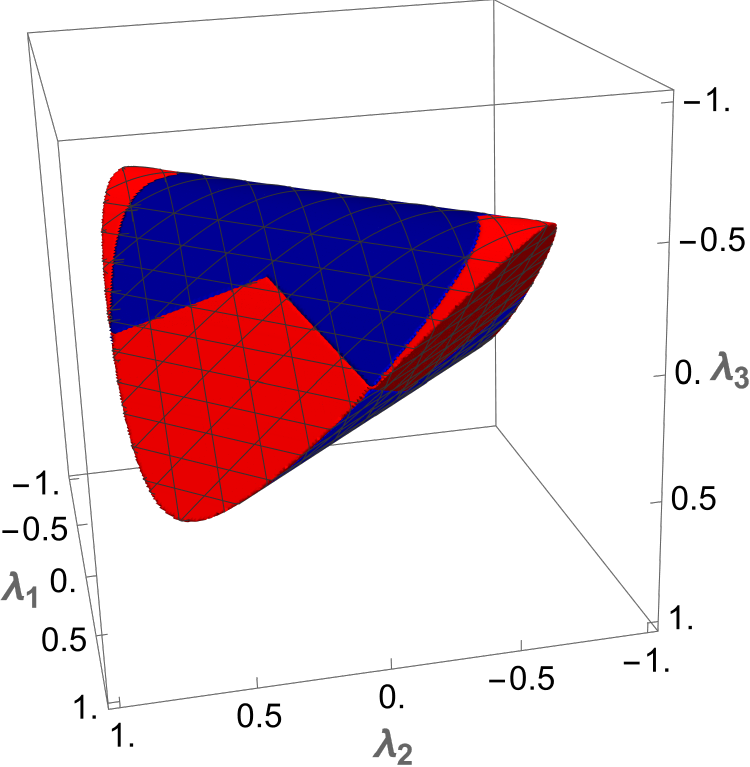}
        \caption{\(t_3 = 1/2\)}
    \end{subfigure}

        \vskip\baselineskip
    
 \begin{subfigure}[b]{0.4\linewidth}
        \centering
        \includegraphics[width=0.8\linewidth]{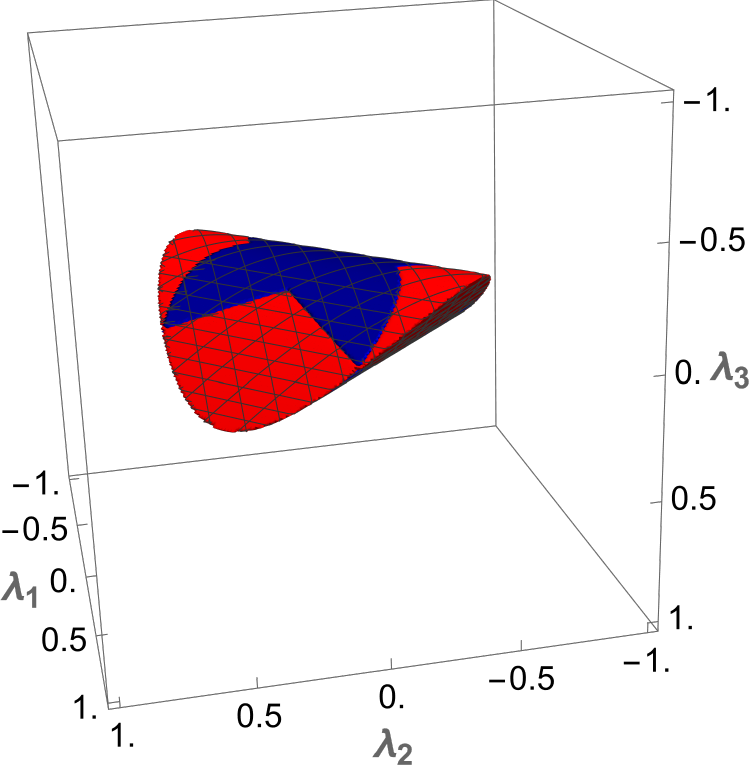}
        \caption{\(t_3 = 3/4\)}
    \end{subfigure}
    \qquad
    \begin{subfigure}[b]{0.4\linewidth}
        \centering
        \includegraphics[width=0.8\linewidth]{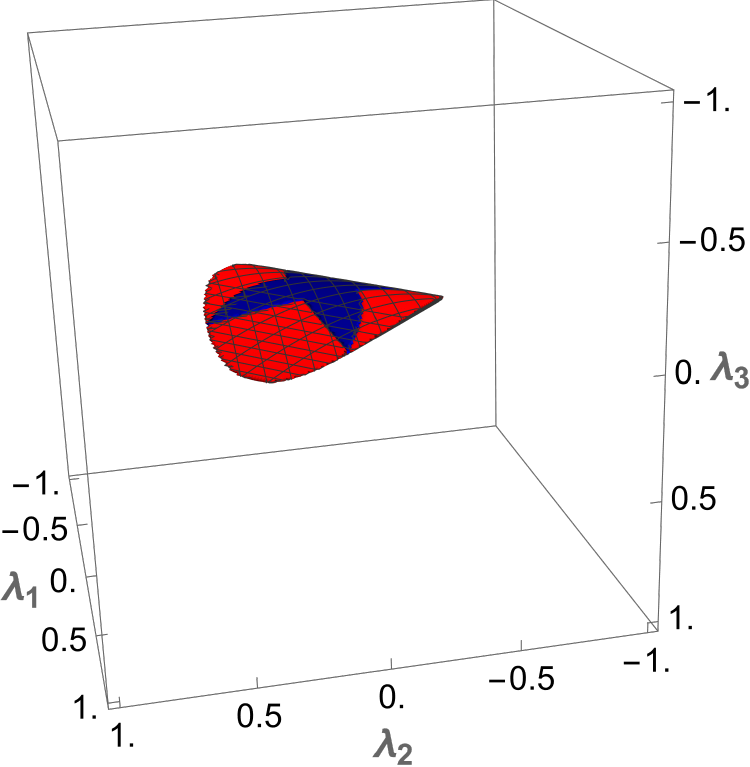}
        \caption{\(t_3 = 7/8\)}
    \end{subfigure}

    \caption{Nonlocality generating channels for different fixed values of $t_3$. Blue and red regions correspond to the channel eigenvalues satisfying conditions (\ref{CH1}) and (\ref{CH2}), respectively.}
    \label{fig:general-case}
\end{figure}

\section{Case analysis}

In what follows, we present examples of nonlocality generating channels among popular qubit channels. Our analysis is divided into two subsections: the Pauli channels, which include all bistochastic qubit channels, and non-unital phase-covariant channels, popular in quantum optics.

\subsection{Pauli channels}

A bistochastic evolution of a qubit is provided by the Pauli channel, which corresponds to the choice $t_k=0$, $k=1,2,3$. Its Kraus representation \cite{King,Landau}
\begin{equation}\label{Pauli}
\Lambda[\rho]=\sum_{\alpha=0}^3 p_\alpha \sigma_\alpha \rho \sigma_\alpha
\end{equation}
reveals a mixed unitary structure with Pauli matrix rotations,
where $p_\alpha$ denotes a probability distribution. Mixed unitary channels, also known as random unitary channels \cite{Scheel} and channels of the random external field type \cite{Alicki,REF}, arise from a unitary evolution that is disrupted by classical errors \cite{TQI,Alicki}. Hence, their noise can be corrected using classical information obtained from measurements on the environment \cite{Gregoratti}. The correspondence between $p_\alpha$ and the channel eigenvalues is given by a simple formula,
\begin{equation}
\lambda_k=2(p_0+p_k)-1, \qquad k=1,2,3.
\end{equation}
Complete positivity of $\Lambda$ is guaranteed by the Fujiwara-Algoet conditions \cite{Fujiwara,King}
\begin{equation}\label{Fuji-2}
|1\pm\lambda_3|\geq|\lambda_1\pm\lambda_2|.
\end{equation}

In general, the nonlocality generating Pauli channels are characterized via $\lambda_1$, $\lambda_2$, $\lambda_3$ such that eqs. (\ref{Fuji-2}) and (\ref{CH1}--\ref{CH2}) hold. These conditions simplify significantly for certain subclasses. Unital qubit channels include many important families of maps \cite{Zyczkowski,Fuchs}:
\begin{enumerate}[label=(\roman*)]
    \item {\it linear channel} with only one non-zero eigenvalue $\lambda_k=\lambda$ such that $|\lambda|\leq 1$;
    \item {\it dephasing channel}, for which one eigenvalue is maximal ($\lambda_k=1$) and the other two are degenerated ($\lambda_\ell=\lambda$ for $\ell\neq k$, $|\lambda|\leq 1$);
    \item {\it depolarizing channel}, where all $\lambda_k=\lambda$ and $-1/3\leq\lambda\leq 1$;
    \item {\it two-Pauli channel}, where $\lambda_1=\lambda_2=\lambda$ and $\lambda_3=2\lambda-1$ for $0\leq\lambda\leq 1$.
\end{enumerate}
Now, it turns out that almost all linear ($|\lambda|<1$) and two-Pauli ($0<\lambda<1$) channels are nonlocality generating. On the other end, we have the dephasing channels, none of which break the CHSH condition. Situated in the middle, there are the depolarizing channels, which generate nonlocality in the finite range $-1/3\leq\lambda\leq 1/\sqrt{2}$. Observe that among each class (i)--(iv), the property of generating nonlocality is controlled via a single parameter $\lambda$.

\subsection{Phase-covariant channels}

Phase-covariant channels are used to describe the dynamics that arises from a combination of energy absorption, energy emission, and pure dephasing \cite{phase-cov-PRL,phase-cov}. They were first considered in describing thermalization and dephasing beyond the Markovian approximation \cite{PC1}. These channels form an important subclass of non-unital qubit channels $\Phi$, which satisfy the covariance property
\begin{equation}\label{cov_def}
\Lambda\big[e^{-i\sigma_3\phi}X e^{i\sigma_3\phi}\big] = e^{-i\sigma_3\phi}\Lambda[X]e^{i\sigma_3\phi}.
\end{equation}
for any input state $X$ and angle $\phi$. On the level of channel eigenvalues, this symmetry property enforces that $\lambda_2=\lambda_1$ \cite{phase-cov,phase-cov-PRL}. The stationary state of $\Lambda$ for $t_3\neq 0$ is no longer a maximally mixed state but instead
\begin{equation}
X_\ast=\frac 12 \left[\mathbb{I}+\frac{t_3}{1-\lambda_3}\sigma_3\right].
\end{equation}
These channels are completely positive if and only if \cite{phase-cov}
\begin{equation}
|\lambda_3|+|t_3|\leq 1,\qquad 4\lambda_1^2+t_3^2\leq(1+\lambda_3)^2.
\end{equation}
Additionally, the necessary and sufficient conditions for a phase-covariant $\Lambda$ to generate nonlocality simplify to
\begin{equation}
\lambda_1^2<\min\{\lambda_3^2,1-\lambda_3^2\}
\quad\mathrm{or}\quad
|\lambda_1|=|\lambda_3|<\frac{1}{\sqrt{2}}
\quad\mathrm{or}\quad
|\lambda_3|<|\lambda_1|<1.
\end{equation}
The above regions are presented visually in Fig.2. Observe that most phase-covariant channels are nonlocality generating, with only two small symmetric regions cut out at the top.

\begin{figure}[H]
    \centering
 \begin{subfigure}[b]{0.4\linewidth}
        \centering
        \includegraphics[width=0.9\linewidth]{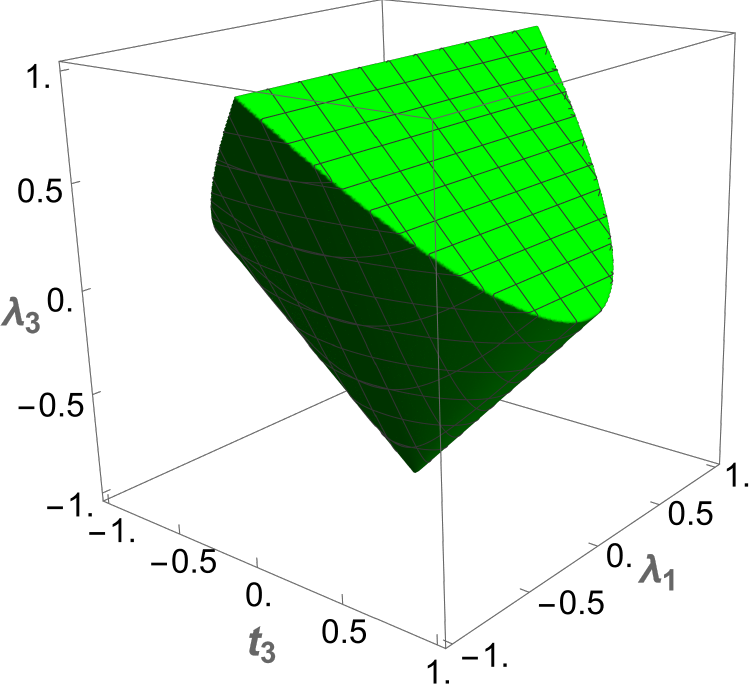}
    \end{subfigure}
    \,
    \begin{subfigure}[b]{0.4\linewidth}
        \centering
        \includegraphics[width=0.9\linewidth]{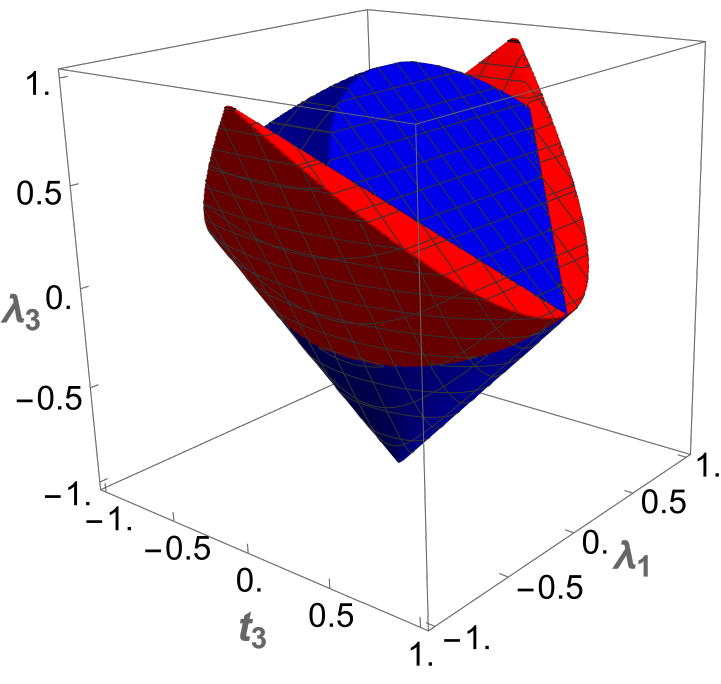}
    \end{subfigure}
    \caption{Region for phase-covariant channels (left) vs. nonlocality generating channels (right).}
\end{figure}

The most popular subclass is represented by generalized amplitude damping channels, which  describe spontaneous emission and absorption of a two-level system in contact with a thermal bath \cite{Myatt,Myatt2}. Generalized amplitude damping corresponds to the choice 
    \begin{equation}
        \lambda_1=\lambda,\qquad \lambda_3=\lambda^2,\qquad t_3=p(1-\lambda^2),
    \end{equation}
    where $|p|\leq 1$ and $|\lambda|\leq 1$. It contains amplitude damping ($p=1$) and inverse amplitude damping ($p=-1$) as special cases. Interestingly, the property of generating nonlocality does not depend on the parameter $p$ but instead is equivalent to
\begin{equation}\label{ADCL}
|\lambda|<1.
\end{equation}
Hence, all generalized amplitude damping channels for which $|\lambda|$ is not maximal correspond to bipartite qubit states that break the CHSH condition. Note that the condition in eq. (\ref{ADCL}) does not depend on the choice of $p$, and it remains the same for both amplitude damping channels and generalized amplitude damping channels. Hence, for the purposes of nonlocality, there is no advantage in allowing for a nonzero temperature of the bath.

In another example, let us consider the generalized shifted depolarizing channels, which are a non-unital analogue of the depolarizing channel \cite{KingNath}. They are characterized by
    \begin{equation}
        \lambda_k=\lambda,\quad k=1,2,3,\qquad t_3=p(1-\lambda),
    \end{equation}
    where $|p|\leq 1$ and $0\leq\lambda\leq 1$.
    This time, nonlocality is generated only by 
\begin{equation}
\lambda<\frac{1}{\sqrt{2}},
\end{equation}
which is less than $2/3$ of all channels from this class.

\section{Conclusions}

In this paper, we investigate the properties of inverse Choi-Jamio{\l}kowski isomorphism, focusing on two-qubit circulant states that satisfy Bell's nonlocality condition. For these states, we construct the corresponding qubit channels and express the CHSH nonlocality criterion in terms of the channel parameters. We show that the region of nonlocality generating channels shrinks the more we deviate from the unital channels. On the example of the Pauli channels and phase-covariant channels, we present how to choose their eigenvalues so that none or almost all maps correspond to nonlocal states.

Our study suggests that nonlocality can be controlled and characterized with fewer parameters by utilizing quantum channels rather than relying on quantum state parameters. This approach has potential applications in every field of quantum information theory where state nonlocality is implemented, such as quantum cryptography \cite{Acin}, quantum games \cite{Cleve}, self-testing of quantum devices \cite{Kaniewski}, and distributed quantum computing \cite{Markham}. An important open question is to generalize our approach to two-qudit or multiqubit scenarios. There is no analogous reformulation of the CHSH Bell inequality in the language of correlation tensors. An interesting approach would be to attempt to reformulate other Bell-type inequalities and to work out their reformulation using quantum channels. \cite{Supic,Salav,Gisin}. It would be also interesting to compare the non-locality generating channels to non-locality breaking channels introduced in ref. \cite{NLBC}. They are defined in analogy to entanglement breaking channels. Namely, a channel is non-locality breaking if and only if its extension acting on any bipartite state produces an output state that does not break the CHSH inequality.

\section{Acknowledgements}

This research was funded in whole or in part by the National Science Centre, Poland, Grant number 2021/43/D/ST2/00102. For the purpose of Open Access, the author has applied a CC-BY public copyright license to any Author Accepted Manuscript (AAM) version arising from this submission.

\bibliographystyle{unsrt}
\bibliography{bibliography}

\end{document}